\documentclass[preprint,12pt]{elsarticle}

\usepackage{graphicx}
\usepackage{amssymb}
\usepackage{amsmath}
\usepackage{physics}
\usepackage{textcomp}
\usepackage{url}
\usepackage[top=1in, bottom=1in, left = 1in, right = 1in]{geometry}
\usepackage[utf8]{inputenc}
\usepackage{nicefrac}
\usepackage{multirow}
\usepackage[export]{adjustbox}
\usepackage{bm}
\usepackage{color, soul}

\usepackage{xcolor}
\usepackage{hyperref}
\usepackage{romannum}
\usepackage{caption}
\usepackage{subcaption}
\usepackage{setspace}
\newcommand{\blue}[1]{{\color{blue}#1}} 

\biboptions{square} 
\usepackage[normalem]{ulem}

\journal{Materialia}

\begin{document}

\pagenumbering{arabic}
\doublespacing
\begin{frontmatter}

\title{Active Learning Sensitivity Analysis of \(\gamma ^\prime\)(L1$_2$) Precipitate Morphology of Ternary Co-Based Superalloys}

\author[NU,CHiMaD]{Whitney Tso}
\ead{whitney.tso@northwestern.edu}

\author[CHiMaD,ANL-2,UMich]{Wenkun Wu}
\ead{wuwenkun@umich.edu}

\author[NU,NUCAPT]{David N. Seidman}
\ead{d-seidman@northwestern.edu}

\author[ANL-2,NAISE,Seagate]{Olle G.~Heinonen}
\ead{olle.heinonen@seagate.com}

\address[NU]{Department of Materials Science and Engineering, Northwestern University, 2220 Campus Drive, Evanston, IL 60208, USA}
\address[CHiMaD]{Center for Hierarchical Materials Design, Northwestern University, 2205 Tech Drive, Suite 1160, Evanston, IL, 60208, USA}
\address[ANL-2]{Materials Science Division, Argonne National Laboratory, 9700 South Cass Avenue, Lemont, IL 60439, USA}
\address[Seagate]{Present and permanent address: Seagate Technology, 7801 Computer Ave., Bloomington, MN 55435}
\address[NUCAPT]{Northwestern University Center Atom-Probe Tomography (NUCAPT), 2220 Campus Drive, Evanston, IL 60208, USA}
\address[NAISE]{Northwestern-Argonne Institute of Science and Engineering, 2205 Tech Drive, Suite 1160, Evanston, Illinois 60208, USA}
\address[UMich]{Department of Materials Science and Engineering, University of Michigan, 2300 Hayward St, Ann Arbor, MI 48109, USA}

\begin{abstract}
To better understand the equilibrium $\gamma^\prime$(L1$_2$) precipitate morphology in Co-based superalloys, a phase field modeling sensitivity analysis is conducted to examine how four phase-field parameters [initial Co concentration ($c_0$), double-well barrier height ($\omega$), gradient energy density coefficient ($\kappa$), and lattice misfit strain ($\epsilon_{\rm misfit}$)] influence the $\gamma^\prime$(L1$_2$) precipitate size and morphology. Gaussian Process Regression (GPR) models are used to fit the sample points and to generate surrogate models for both precipitate size and morphology. In an Active Learning approach, a Bayesian Optimization algorithm is coupled with the GPR models to suggest new sample points to calculate and efficiently update the models based on a reduction of uncertainty. The algorithm has a user-defined objective, which controls the balance between exploration and exploitation for new suggested points. Our methodology provides a qualitative and quantitative relationship between the $\gamma^\prime$(L1$_2$) precipitate size and morphology and the four phase-field parameters, and concludes that the most sensitive phase-field parameter for precipitate size and morphology is the initial Co concentration ($c_0$) and the double-well barrier height ($\omega$), respectively. We note that the GPR model for precipitate morphology required adding a noise tolerance in order to avoid overfitting due to irregularities in some of the simulated equilibrium $\gamma^\prime$(L1$_2$) precipitate morphology.
\end{abstract}


\end{frontmatter}

\section{Introduction}
Ni-based superalloys comprise one of the most important alloy systems for high-temperature applications, such as gas turbine jet engines, land-based natural gas-fired engines for producing electricity, and nuclear reactor components. \cite{xia2020} The exceptional mechanical properties exhibited by Ni-based superalloys originate in their unique two-phase microstructure with a $\gamma^\prime$(L1$_2$)-precipitate phase in a disordered face-centered cubic (f.c.c) $\gamma$-matrix phase. The $\gamma^\prime$(L1$_2$)-precipitate phase is elastically anisotropic and preferentially grows along the [001] crystallographic directions, which leads to formation of cuboidal $\gamma^\prime$(L1$_2$)-precipitates, which provide significant precipitate strengthening and creep resistance. 

Recently, Co-based superalloys have attracted much attention as potential replacements for conventional Ni-based superalloys used in high-temperature applications.  Sato and co-workers \cite{sato2006} demonstrated that the Co-Al-W ternary alloy system exhibits a thermodynamically stable $\gamma$(f.c.c.)/$\gamma^\prime$(L1$_2$) two-phase region at 900$^{\circ}$C. Furthermore, Co-based superalloys show promise for having higher operating temperatures and better corrosion resistance than Ni-based superalloys. \cite{sato2006,gao2019} Despite having a similar $\gamma$(f.c.c.)/$\gamma^\prime$(L1$_2$) two-phase region to Ni-based superalloys, the chemistry is quite different, and the existing body of knowledge of Ni-based superalloys often cannot be directly transferred to Co-based superalloys. Current research has been focused on better understanding how different factors influence the equilibrium microstructure, specifically the effects of alloying elements \cite{mottura2012,suzuki2008,chen2010,chung2020} and external stresses \cite{wang2019,chung2020_raft}. 

With increasing computational capabilities, there is a paradigm shift to use more computational tools to explore the microstructural evolution \cite{mottura2012,wang2019,chen2010} and to use artificial intelligence to aid in materials design. \cite{liu2015,hart2021} Considering how time-consuming and expensive it is to perform large numbers of experiments, analysis through computational modeling provides a more cost-effective alternative that allows for every parameter to be systematically examined. For example, phase-field modeling is an analytical method, which simulates microstructural evolution at the mesoscale \cite{moelans2007,thompson1999}.

Machine learning is an extremely versatile tool that can provide solutions to a wide range of problems with the help of artificial intelligence. For materials design, a surrogate model can be fitted with existing data, and it can be used to identify optimal compositions and processing parameters to achieve the desired properties. This allows for more efficient materials design compared to the traditional trial-and-error method \cite{liu2015,gubernatis2018}. Additionally, sensitivity analysis can also improve the materials design process by increasing our understanding of material systems. Particularly for alloy systems, we can determine how specific input parameters, such as composition and processing steps, can influence the equilibrium microstructure of the final product. Information gained from sensitivity analysis can then be used to fine tune and design for the desired microstructures and properties by modifying the input parameters.

In a previous Co-based superalloys phase-field study by Wu \emph{et al.}\cite{wu2022}, a sensitivity analysis was conducted to better understand how different phase-field parameters influence the equilibrium $\gamma^\prime$(L1$_2$) precipitate morphology in a ternary Co-0.1Al-(0.9-x)W system. The sensitivity was then expressed as a standard bi-linear equation that was obtained by fitting sample inputs selected with a Latin hypercube. In the present study, our goal is to use a data-driven approach to fit a more sophisticated surrogate model for the relationship between microstructure and input parameters. The surrogate model selected in this work is based on a Gaussian Process Regression (GPR) model. A regression-based model is used because the \(\gamma^\prime\)(L1$_2$) precipitate morphology predictions are a continuous variable; the GPR model is also an example of a machine learning model that is suitable for situations where relatively small data sets are available,\cite{schulz2018} as is the case in the present work. 

Active learning is another important aspect in machine learning that iteratively updates an algorithm or model based on an objective. Bayesian Optimization (also known as Optimal Experimental Design) is an example of active learning that is used in the present research. In this approach optimal new data points are suggested based on a user-defined objective and the surrogate model is updated based on previous outputs and outputs from the new data points. This specific algorithm was selected in our case because it is ideal for situations where there is a low dimensionality in the exploration space \cite{frazier2015}.

\section{Method}
The phase-field model used in this article is based on the model formulated by Wu \emph{et al.}\cite{wu2022}. In this model, the concentration field $c$ is simplified to only describing the mole fraction of Co in the Co-0.1Al-(0.9-x)W ternary alloy system, where x is between 0.808 and 0.86. It has been demonstrated that the Al concentration is close to a constant across the \(\gamma\)(f.c.c.) and $\gamma^\prime$(L1$_2$) phases based on experimental data \cite{bocchini2013}. In addition, W is a slow diffusing element in Co, which is why the W diffusion is assumed to be negligible in this phase-field model. The phase field evolution is driven by the Cahn-Hilliard equation \cite{cahn1958} for a concentration field $c$: 
\begin{equation}\label{Cahn-Hilliard_eqn}
    \frac{\partial c}{\partial t}=\nabla\cdot (M\nabla\frac{\partial F}{\partial c}),
\end{equation}
and the Allen-Cahn equation \cite{allen1972} for the order parameter (\(\eta\)): 
\begin{equation}\label{Allen-Cahn_eqn}
    \frac{\partial \eta}{\partial t}=-L\frac{\partial F}{\partial \eta}.
\end{equation}
where \(F\) is the Helmholtz free energy density of the system which is comprised of three components: (i) local chemical free energy density, (ii) gradient energy density, and (iii) elastic energy density; \(t\) is time, \(M\) is the mobility coefficient, and \(L\) is a kinetic coefficient that sets a scale for the time-evolution of the non-conserved order parameter $\eta$ which is used to differentiate between the \(\gamma\)(f.c.c.)- and \(\gamma'\)(L1$_2$)- phases. The chemical free energy consists of an interpolation of the composition-dependent bulk chemical free energy density of each phase, and a double-well potential for the phase-field order parameter, $\eta$. The local chemical free energy density is described by
\begin{equation}\label{chem-energy_eqn}
    f_{\rm chem}(c,\eta)=f_{\gamma}(c_{\gamma})(1-h(\eta))+f_{\gamma'}(c_{\gamma^\prime})h(\eta)+\omega g(\eta),
\end{equation}
where \(f_{\gamma}\) and \(f_{\gamma^\prime}\) are the free energy densities of the \(\gamma\)(f.c.c.)- and \(\gamma'\)(L1$_2$)- phases, \(c_{\gamma}\) and \(c_{\gamma'}\) are the compositions of the \(\gamma\)(f.c.c.)- and \(\gamma'\)(L1$_2$)- phases, \(h(\eta)\) is an interpolation function, \(\omega\) is a double-well potential barrier height, and \(g(\eta)\) is a double-well potential. Further details concerning the phase-field model are in a prior article by Wu \emph{et al.}\cite{wu2022}. 

The gradient energy density (\(f_{\rm grad}\)) and elastic energy density (\(f_{\rm elas}\)) are described by the following standard equations:
\begin{equation}\label{grad-energy_eqn}
    f_{\rm grad}=\frac{\kappa}{2}(\nabla \eta)^2
\end{equation}
\begin{equation}\label{elas-energy_eqn}
    f_{\rm elas}=\frac{1}{2}\sigma_{ij}\epsilon_{ij}^{\rm elas}
\end{equation}
where \(\kappa\) is the gradient energy density coefficient, \(\sigma_{ij}\) is the elastic stress, and \(\epsilon_{ij}^{\rm elas}\) is the elastic strain. All the phase-field parameters are derived from a combination of experimental and modeling data: for details, see Ref.~\cite{wu2022}. The double-well potential height ($\omega$) describes an activation energy for transformation between the $\gamma$(f.c.c.)- and $\gamma'$(L1$_2$)- phases, represented by the phase-field values $\eta$ = 0 and $\eta$ = 1, respectively. The double-well potential ($\omega$) and gradient energy density coefficient ($\kappa$) are directly related to  the interfacial width between the $\gamma$(f.c.c.)- and $\gamma'$(L1$_2$)- phases and the interfacial free energy. 
The lattice misfit strain ($\epsilon_{\rm misfit}$) was constrained to a range with stable $\gamma'$(L1$_2$) precipitates, where the precipitates neither shrink nor grow larger than the phase-field simulation cell. For example, if $\epsilon_{\rm misfit}$ is too high, the elastic energy required to maintain a $\gamma'$(L1$_2$) precipitate will be too large, and thus lead to the shrinkage and disappearance of $\gamma'$ precipitate. For the initial sample set, the range for the kinetic coefficient $L$ was carefully selected to ensure the phase-field simulation is in a physically reasonable region in which the microstructural evolution is dominated by diffusion of the concentration field $c$. The kinetic  coefficient $L$ controls the time-evolution of the order parameter $\eta$, therefore \(\eta\) lags the concentration field for small values of $L$. Alternatively, for large values of $L$ the phase field evolution is much more rapid than that of the concentration field \(\eta\) \cite{wu2022}. 
Based on prior work \cite{wu2022}, we know that the \(R_0\) has an insignificant effect on the equilibrium precipitate size and morphology. The same is true for $L$ as long as it is maintained in a range in which the morphological evolution is dominated by concentration field diffusion, as discussed earlier. Therefore, we fixed \(R_0\) and $L$ to 75 and 100, respectively, to reduce the parameter space dimensions. The parameter space is listed in Table~ \ref{search_space_table}, which was constrained to normalized input parameters in physically reasonable ranges. 


Six phase field input parameters were considered for the sensitivity analysis: initial Co concentration (\(c_0\)), double-well potential barrier height (\(\omega\)), gradient energy density coefficient (\(\kappa\)), lattice parameter misfit strain (\(\epsilon_{\rm misfit}\)) 
, initial \(\gamma'\)(L1$_2$) precipitate size (\(R_0\)), and the kinetic coefficient $L$ in the Allen-Cahn equation. The initial state of the simulation cell included one spherical \(\gamma'\)(L1$_2$) precipitate centered at the origin with a specified initial size. The total free energy of the system was then minimized by evolving the coupled Cahn-Hilliard and Allen-Cahn equations to reach a local equilibrium state. 

An initial ten samples with varying six parameters (\(c_0, \omega, \kappa, \epsilon_{\rm misfit}, R_0, L\)) were selected using a Latin hypercube, which are listed in Table~\ref{samples_table} \cite{wu2022}. Some samples exhibited irregular precipitate morphologies or were unstable and were then treated as outliers and removed from the dataset. The samples that were removed are samples 2, 3, 5, 8, and 10. Five initial samples were fitted to two preliminary GPR models for equilibrium precipitate size and morphology, respectively. Samples 11 - 20 in Table~\ref{samples_table} are additional samples suggested using Bayesian Optimization algorithm. All input parameters were normalized based on the constrained parameter space. The mean precipitate size (\(r_{\rm mean}\)) was defined to be the average distance between the centers of the precipitates to the interface with \(0.3 < \eta < 0.7\). Precipitate morphology (\(k_{\rm cubic}\)) was calculated with 
\begin{equation}\label{k_cubic_eqn}
    k_{\rm cubic}=1-|\frac{r_{\rm max}}{\sqrt{3}r_{\rm min}}-1|,
\end{equation}
which is a measure of how cubic the precipitate is; \(r_{\rm max}\) and \(r_{\rm min}\) represents the maximum and minimum distances from the center to the interface of the precipitate respectively. A perfect cube yields unity, whereas a more rounded precipitate morphology will have a lower \(k_{\rm cubic}\) value. 

\begin{table}
\begin{center}
    \caption{\label{samples_table}List of samples with a solid line that demarcates Latin hypercube samples and optimized samples\cite{wu2022}. 
    The quantity $c_0$ is the initial mole fraction of Co and is dimensionless, $\omega$ is the double-well barrier height with units of G$\cross$J/m$^3$ (or 10$^9$ $\cross$ J/m$^3$), $\kappa$ is the gradient energy density coefficient with units of n$\cross$J/m (or 10$^{-9}$ $\cross$ J/m), $\epsilon_{\rm misfit}$ is the lattice parameter misfit strain and is dimensionless, $R_0$ is the initial precipitate size with units of nm, and $L$ is the kinetic coefficient in units of 10$^{-45}$m$^3$s$\cdot$J). $^*$ marks the reference sample.}
    \begin{tabular}{ c c c c c c c }
    \hline
    Sample & \(c_0\) & \(\omega\) & \(\kappa\) & \(\epsilon_{\rm misfit}\) & \(R_0\) & L \\
    \hline
    1 & 0.849 & 0.128 & 1.366 & \(0.545\%\) & 95.93 & 55.37 \\
    2 & 0.803 & 0.161 & 1.180 & \(0.324\%\) & 25.38 & 87.21 \\
    3 & 0.834 & 0.297 & 0.699 & \(0.852\%\) & 184.6 & 119.0 \\
    4 & 0.843 & 0.191 & 0.762 & \(0.265\%\) & 53.12 & 184.4 \\
    5 & 0.860 & 0.314 & 0.607 & \(0.990\%\) & 120.0 & 66.92 \\
    6 & 0.841 & 0.209 & 1.007 & \(0.757\%\) & 175.6 & 76.32 \\
    7 & 0.823 & 0.111 & 0.535 & \(0.424\%\) & 67.22 & 60.26 \\
    8 & 0.817 & 0.141 & 1.834 & \(0.474\%\) & 77.51 & 114.0 \\
    9 & 0.829 & 0.096 & 1.592 & \(0.373\%\) & 45.22 & 155.4 \\
    10 & 0.808 & 0.251 & 0.995 & \(0.595\%\) & 34.66 & 150.0 \\
    \hline
    11 & 0.822 & 0.243 & 1.007 & \(0.753\%\) & 75.00 & 100.0 \\
    12 & 0.827 & 0.185 & 1.608 & \(0.620\%\) & 75.00 & 100.0 \\
    13$^*$ & 0.831 & 0.256 & 1.258 & \(0.620\%\) & 75.00 & 100.0 \\
    14 & 0.826 & 0.203 & 1.583 & \(0.634\%\) & 75.00 & 100.0 \\
    15 & 0.818 & 0.234 & 1.733 & \(0.723\%\) & 75.00 & 100.0 \\
    16 & 0.823 & 0.287 & 1.408 & \(0.708\%\) & 75.00 & 100.0 \\
    17 & 0.836 & 0.243 & 1.157 & \(0.575\%\) & 75.00 & 100.0 \\
    18 & 0.819 & 0.222 & 1.382 & \(0.646\%\) & 75.00 & 100.0 \\
    19 & 0.819 & 0.199 & 1.446 & \(0.646\%\) & 75.00 & 100.0 \\
    20 & 0.819 & 0.211 & 1.382 & \(0.608\%\) & 75.00 & 100.0 \\
    \end{tabular}
    
\end{center}
\end{table}

\begin{table} 
\begin{center} 
    \caption{\label{search_space_table}The constrained parameter space for Bayesian Optimization.}
    \begin{tabular}{ c c c c c }
    \hline
      & \(c_0\) & \(\omega\) & \(\kappa\) & \(\epsilon_{\rm misfit}\) \\
    \hline
    min & 0.808 & 0.096 & 0.607 & \(0.265\%\) \\
    max & 0.86 & 0.314 & 1.834 & \(0.757\%\) \\
    \end{tabular}
\end{center}
\end{table}

The Bayesian Optimization (BO) approach developed in this article consists of three main steps. The first step involves fitting the GPR models using the initial samples as training data. The second step consists of assessing uncertainty quantification and selecting a new point in parameter space, balancing exploitation and exploration for the suggested sample points, which is controlled by maximizing the acquisition function. Typically, the goal of the acquisition function is to reduce the uncertainty in the surrogate model \cite{frazier2018}. The expected improvement acquisition function ($E(x)$) is commonly used in Bayesian Optimization \cite{frazier2018}, which is computed with 
\begin{equation}\label{EI_eqn}
    E(x)=(\mu(x)-f(x^+)-\xi)\Psi(\frac{\mu(x)-f(x^+)-\xi}{\sigma(x)})+\sigma(x)\phi(\frac{\mu(x)-f(x^+)-\xi}{\sigma(x)}),
\end{equation}
where \(E(x)\) is the expected improvement acquisition function, \(x\) represents a set of inputs, \(\mu\) is the predicted value of the given inputs using the surrogate model, \(f(x^+)\) is the best sample based on the user-defined objective, \(\xi\) is the exploration term that is set equal to 0.01,  
\(\Psi\) is the cumulative distribution function, \(\phi\) is the probability density function of a standard Gaussian distribution, and \(\sigma\) is the standard deviation of the surrogate model.


Exploitation places an emphasis on sample points that are optimally based on a specific objective, whereas exploration focuses on examining larger regions of the parameter space. In this study, the objective is to have an accurate surrogate model that gives the relationship between phase-field parameters and the $\gamma^\prime$(L1$_2$) precipitate size and morphology. The exploration term ($\xi$) controls the balance between exploitation and exploration. There is no exploration when $\xi = 0$; whereas a larger $\xi$ term leads to more exploration of the parameter space. \cite{jasrasaria2018} A relatively small exploration term ($\xi = 0.01$) is selected, so that the BO algorithm mainly focuses on suggesting new sample points, which reduces the uncertainty of the model, with a small degree of exploration to keep away from the edge of the parameter space. Computing new sample points at the edge of the parameter space is less likely to improve the accuracy of the surrogate model and is less important for interpreting the surface response. 

The parameter space is listed in Table~ \ref{search_space_table}, which is constrained to normalize the input parameters in a range where there are thermodynamically stable $\gamma'$(L1$_2$) precipitates based on experimental and simulated data. The third and final step in the BO scheme is to perform phase-field simulations for the suggested sample points and to iteratively update the surrogate models. 

\section{Results}
\subsection{Precipitate Size}

To visualize the GPR models in a 3-D plot, the precipitate size or morphology are plotted with two phase-field parameters, while the other two phase-field parameters are fixed at constant values. The phase-field parameters can be fixed at any values within the parameter space. Sample 13 from Table~\ref{samples_table} is selected as the reference sample to generate the plots displayed in this article because the sample is relatively close to the center of the parameter space. Additionally, the phase-field results from Sample 13 can be used to verify the accuracy of the GPR models. Samples 11 and 18 are used as reference samples to plot the same GPR models in the Appendix to demonstrate that the choice of reference point is arbitrary.

\begin{figure}
\begin{subfigure}{0.45\textwidth}
    \centering
    \includegraphics[width=\textwidth]{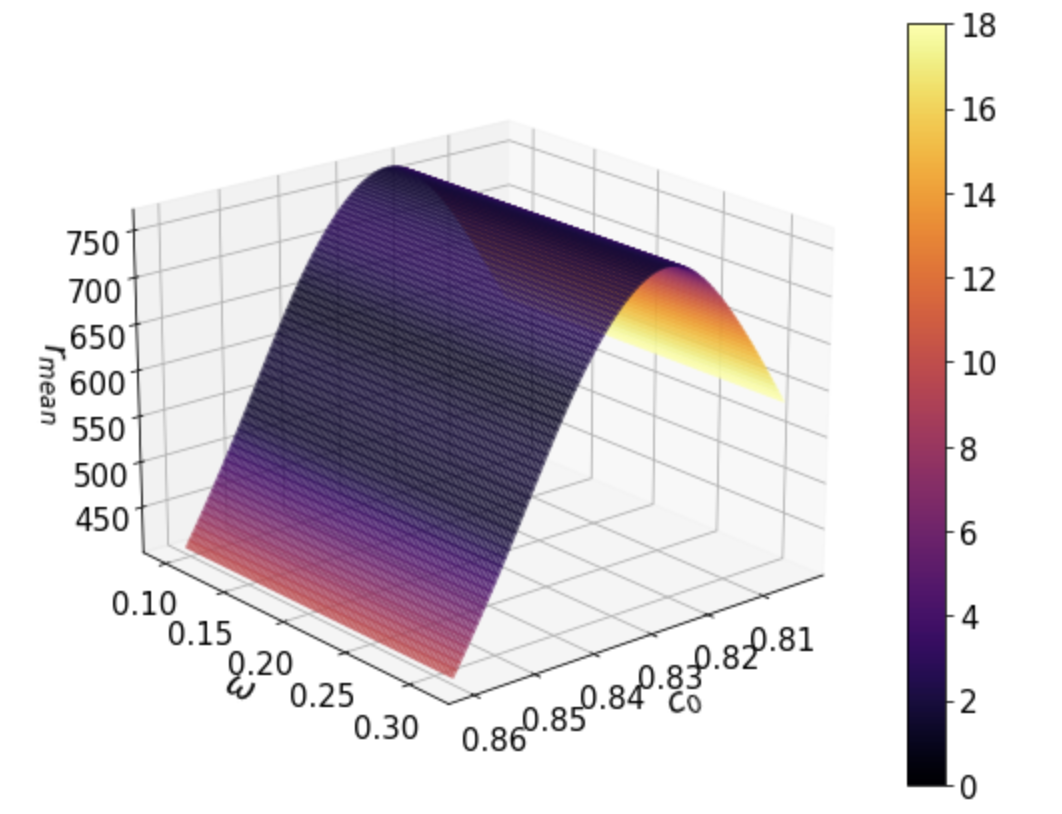}
    \caption{}
    \label{fig:r-c-w_initial}
\end{subfigure}
\begin{subfigure}{0.45\textwidth}
    \centering
    \includegraphics[width=\textwidth]{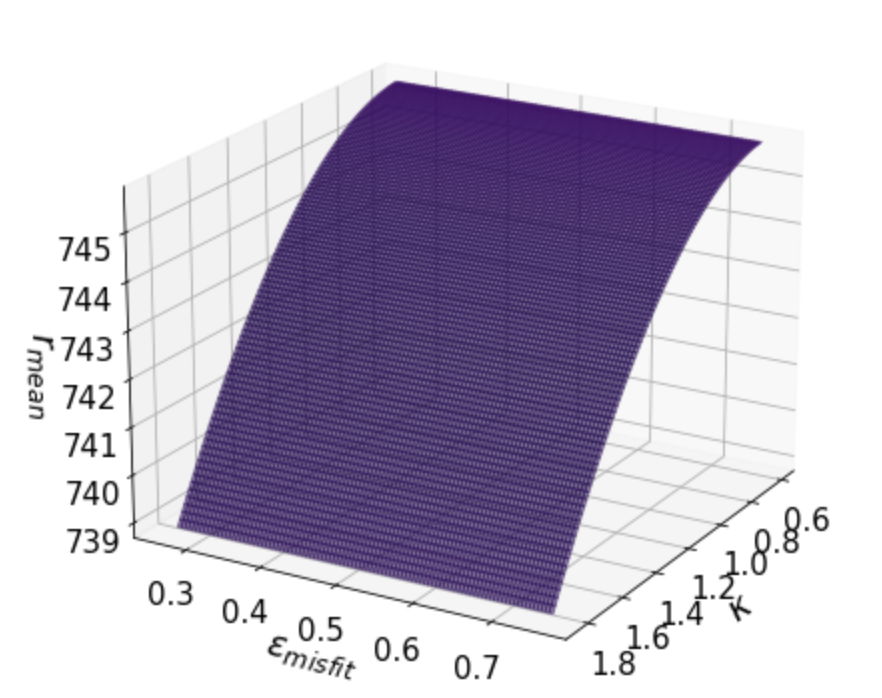}
    \caption{}
    \label{fig:r-k-eps-initial}
\end{subfigure}
\caption{Initial GPR surrogate model of the precipitate size fitted with five samples. The surface plot colors and scale bar in the middle represents the standard deviation of the model, which is shared between the two plots. \(r_{\rm mean}\) and the standard deviation are in nm and \(c_0\) and \(\epsilon_{\rm misfit}\) are dimensionless, \(\omega\) is G$\cross$J/m$^3$ (or 10$^9$ $\cross$ J/m$^3$), and \(\kappa\) is n$\cross$J/m (or 10$^{-9}$ $\cross$ J/m). (a) is plotted by taking $\kappa$ = 1.258 n$\cross$J/m (or 10$^{-9}$ $\cross$ J/m) and $\epsilon_{\rm misfit}$ = 0.62\%. (b) is plotted by taking $c_0$ = 0.831 and $\omega$ = 0.256 G$\cross$J/m$^3$ (or 10$^9$ $\cross$ J/m$^3$). The two plots employ different z-axis scaling. In comparison, the initial Co concentration \(c_0\) is the most sensitive phase-field parameter. }
\label{fig:r_initial}
\end{figure}

\begin{figure}
\begin{subfigure}{0.45\textwidth}
    \centering
    \includegraphics[width=\textwidth]{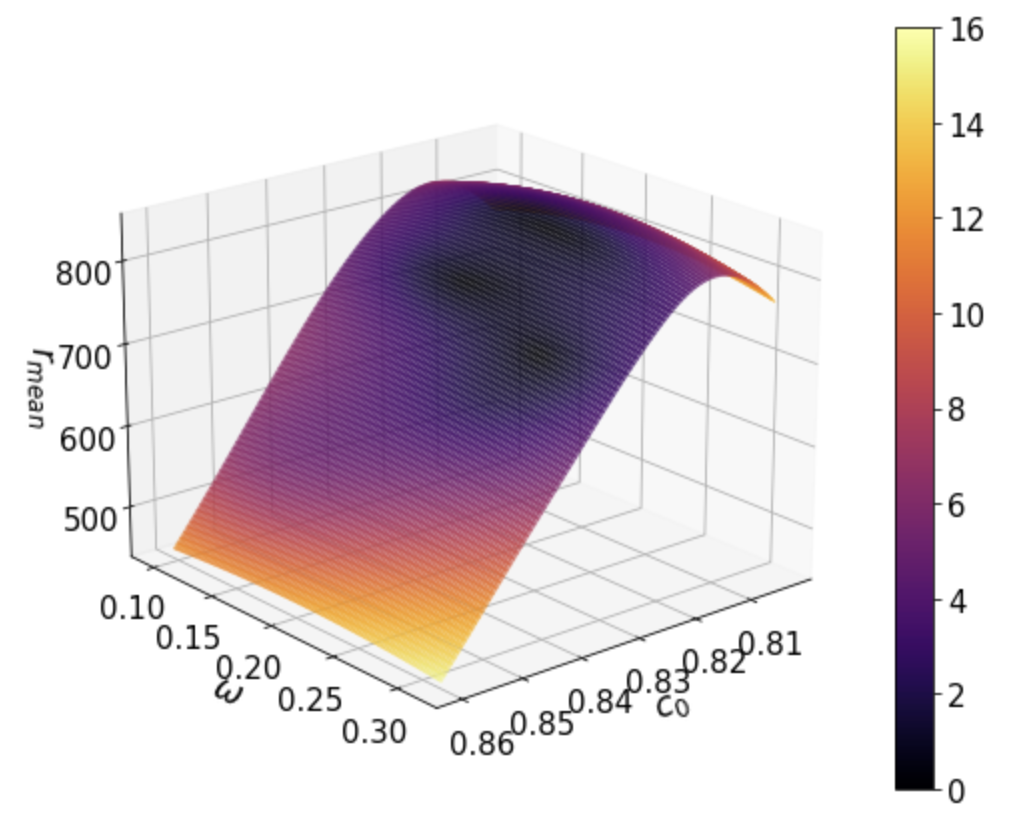}
    \caption{}
    \label{fig:r-c-w_final}
\end{subfigure}
\begin{subfigure}{0.45\textwidth}
    \centering
    \includegraphics[width=\textwidth]{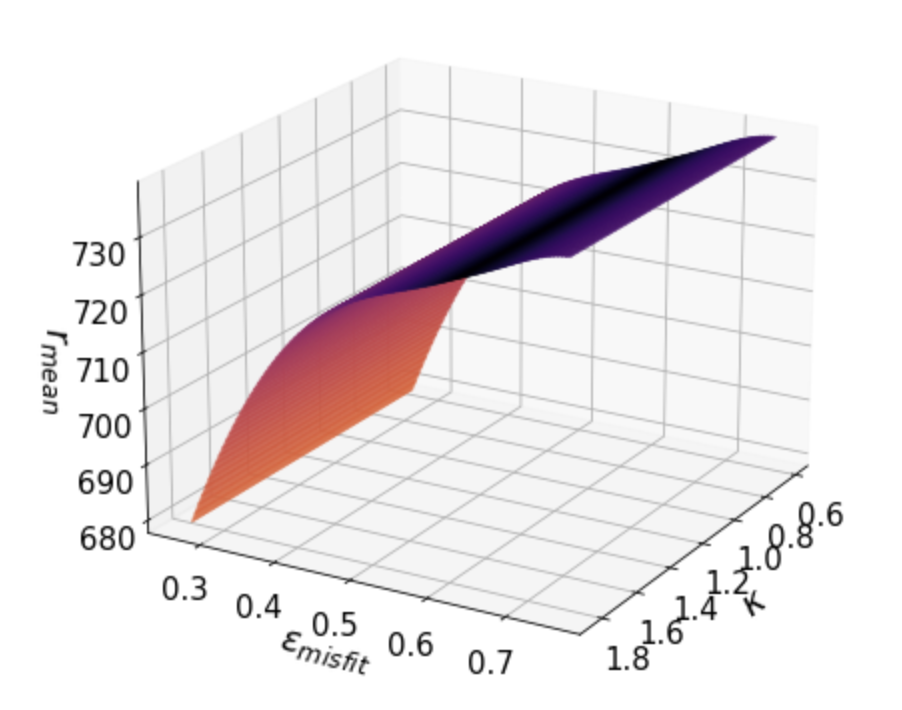}
    \caption{}
    \label{fig:r-k-eps_final}
\end{subfigure}
\caption{Final GPR surrogate model of the precipitate size fitted with all samples. The surface plot colors and scale bar in the middle represents the standard deviation of the model, which is shared between the two plots. \(r_{\rm mean}\) and the standard deviation are in nm and \(c_0\) and \(\epsilon_{\rm misfit}\) are dimensionless, \(\omega\) is G$\cross$J/m$^3$ (or 10$^9$ $\cross$ J/m$^3$), and \(\kappa\) is n$\cross$J/m (or 10$^{-9}$ $\cross$ J/m). (a) is plotted by fixing $\kappa$ = 1.258 n$\cross$J/m (or 10$^{-9}$ $\cross$ J/m) and $\epsilon_{\rm misfit}$ = 0.62\%. (b) is plotted by fixing $c_0$ = 0.831 and $\omega$ = 0.256 G$\cross$J/m$^3$ (or 10$^9$ $\cross$ J/m$^3$). The plots (a) and (b) have different z-axis scaling. In comparison, the initial Co concentration \(c_0\) is the most sensitive phase-field parameter.}
\label{fig:r_final}
\end{figure}

\begin{figure}
    \centering
    \includegraphics[width=0.7\textwidth]{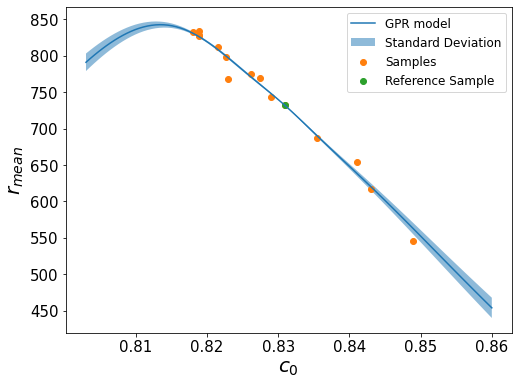}
    \caption{Final GPR model of precipitate size vs initial concentration. The green dot is the reference sample and orange dots are all the samples. The blue solid line marks the GPR model's prediction and the light blue shade displays the standard deviation. This displays a strong dependence of $\gamma^\prime$(L1$_2$) precipitate size to the mole fraction of Co.}
    \label{fig:r-c}
\end{figure}

A preliminary GPR model was fitted with the initial five samples selected with Latin hypercube, Fig.~\ref{fig:r_initial}, where we observe a strong dependence of precipitate size on the initial concentration.  The two plots in Fig.~\ref{fig:r_initial} are generated by holding two out of four phase-field parameters constant to the reference sample, which is Sample 13 in Table \ref{samples_table}.  
Figure~\ref{fig:r-c-w_initial} plots \(r_{\rm mean}\) with respect to \(c_0\) and \(\omega\), while keeping \(\kappa\) and \(\epsilon_{\rm misfit}\) constant at 1.258 n-J/m (or 10$^{-9}$ $\cross$ J/m) and 0.62\% respectively. Figure~\ref{fig:r-k-eps-initial} plots \(r_{\rm mean}\) with respect to \(\kappa\) and \(\epsilon_{\rm misfit}\), while maintaining \(c_0\) and \(\omega\) at 0.831 and 0.256 n-J/m (or 10$^{-9}$ $\cross$ J/m)\(^3\) respectively. 
The standard deviation of the model is displayed employing colors in the 3-D surface plots. The initial model shows that there is higher uncertainty at a small initial concentration. 
Figure~\ref{fig:r_final} displays the GPR model after the final iteration with the GPR model updated with the new samples suggested with BO. The high uncertainty regions are now limited to the boundaries of the exploration space. 
This is desirable because it indicates greater accuracy in the region of interest, which is in the middle of the search space. Like the initial model, the final model indicates that the precipitate size is most sensitive to the initial Co concentration. The other three phase-field parameters ($\omega$, $\kappa$, and $\epsilon_{\rm misfit}$) have little influence on the precipitate size. Figure~\ref{fig:r-k-eps_final} is plotted with a smaller z-axis range compared to Figure~\ref{fig:r-c-w_final}, which demonstrates that $\kappa$ and $\epsilon_{\rm misfit}$ have a minor influence on the precipitate size. While both the initial and final models predicted similar trends between the precipitate size and the four phase-field parameters, the predicted values are significantly changed after the addition of new samples. 
Figure~\ref{fig:r-c} compresses the multi-dimensional model to a 2D plot that demonstrates the relationship between the most sensitive parameter, \(c_0\), with precipitate size.  

\subsection{Precipitate Morphology}
\begin{figure}
\begin{subfigure}{0.45\textwidth}
    \centering
    \includegraphics[width=\textwidth]{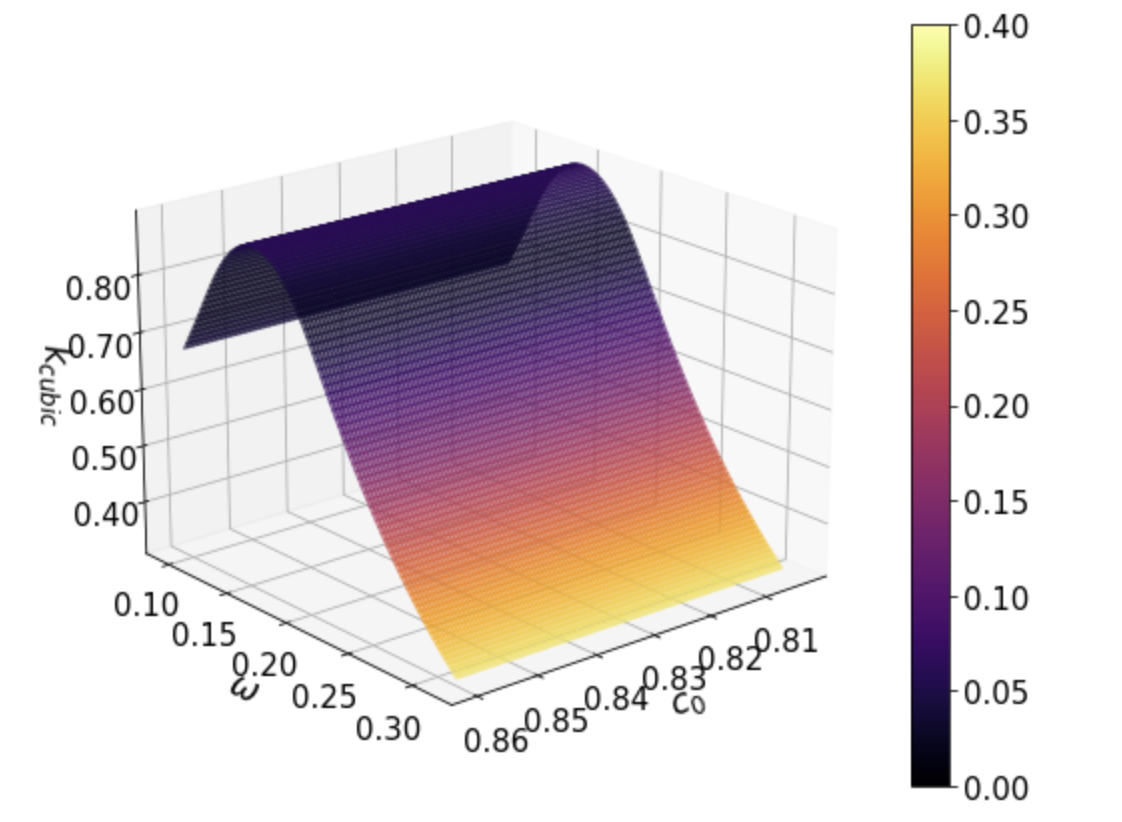}
    \caption{}
    \label{fig:k-c-w_initial}
\end{subfigure}
\begin{subfigure}{.45\textwidth}
    \centering
    \includegraphics[width=\textwidth]{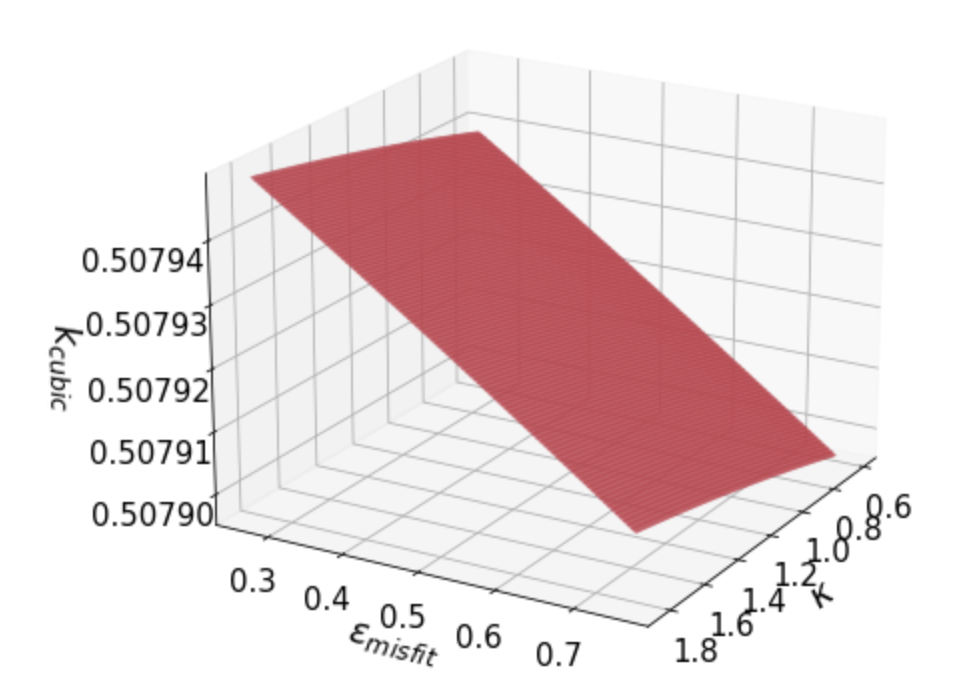}
    \caption{}
    \label{fig:k-k-eps_initial}
\end{subfigure}
\caption{Initial GPR surrogate model of precipitate morphology fitted with five samples. The surface plot color and the scale bar in the middle represents the standard deviation of the model, which is shared between the two plots. \(k_{\rm cubic}\), standard deviation, \(c_0\), and \(\epsilon_{misfit}\) are dimensionless, \(\omega\) is G$\cross$J/m$^3$ (or 10$^9$ $\cross$ J/m$^3$), and \(\kappa\) is n$\cross$J/m (or 10$^{-9}$ $\cross$ J/m). (a) is plotted by taking $\kappa$ = 1.258 n$\cross$J/m (or 10$^{-9}$ $\cross$ J/m) and $\epsilon_{\rm misfit}$ = 0.62\%. (b) is plotted by keeping $c_0$ = 0.831 and $\omega$ = 0.256 G$\cross$J/m$^3$ (or 10$^9$ $\cross$ J/m$^3$). These plots have different z-axis scaling. In comparison, the double-well barrier height \(\omega\) is the most sensitive phase-field parameter.}
\label{fig:k_initial}
\end{figure}

\begin{figure}
\begin{subfigure}{0.45\textwidth}
    \centering
    \includegraphics[width=\textwidth]{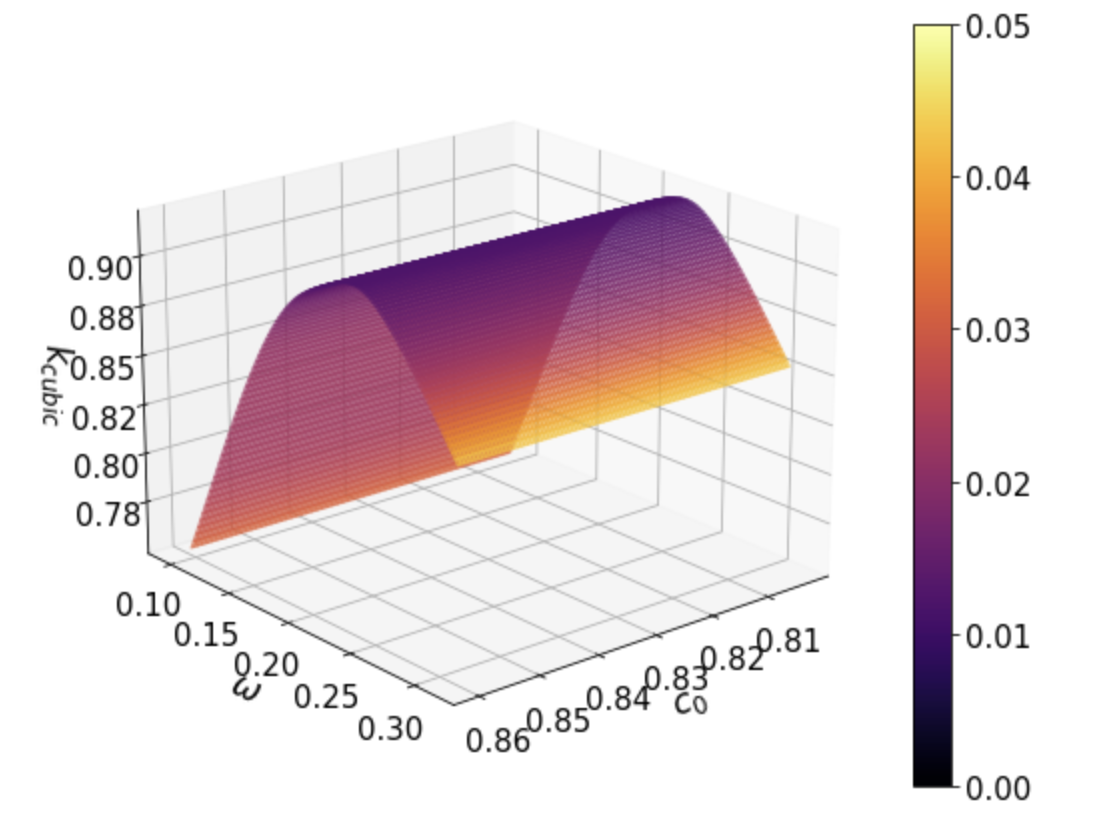}
    \caption{}
    \label{fig:k-c-w_final}
\end{subfigure}
\begin{subfigure}{.45\textwidth}
    \centering
    \includegraphics[width=\textwidth]{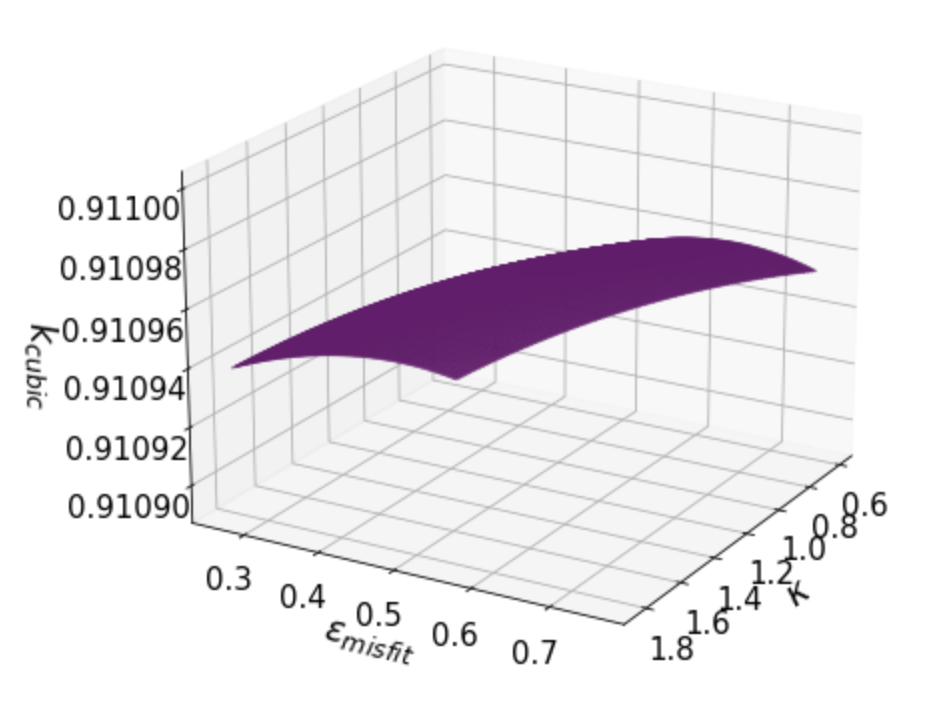}
    \caption{}
    \label{fig:k-k-eps_final}
\end{subfigure}
\caption{Final GPR surrogate model of precipitate morphology fitted with all samples. The surface plot color and the scale bar between them represents the standard deviation of the model, which is shared between the two plots. \(k_{\rm cubic}\), standard deviation, \(c_0\), and \(\epsilon_{\rm misfit}\) are dimensionless, \(\omega\) is G$\cross$J/m$^3$ (or 10$^9$ $\cross$ J/m$^3$), and \(\kappa\) is n$\cross$J/m (or 10$^{-9}$ $\cross$ J/m). (a) is plotted by fixing $\kappa$ = 1.258 n$\cross$J/m (or 10$^{-9}$ $\cross$ J/m) and $\epsilon_{\rm misfit}$ = 0.62\%. (b) is plotted by fixing $c_0$ = 0.831 and $\omega$ = 0.256 G$\cross$J/m$^3$ (or 10$^9$ $\cross$ J/m$^3$). The plots have different z-axis scaling. In comparison, the double-well barrier height \(\omega\) is the most sensitive phase-field parameter.}
\label{fig:k_final}
\end{figure}

Like the precipitate size case, a preliminary precipitate morphology GPR model is fitted with the initial five samples, Fig.~\ref{fig:k_initial}. The two plots in Fig.~\ref{fig:k_initial} are generated by holding the two phase-field parameters constant with respect to the reference sample, which is Sample 13 in Table~\ref{samples_table}. Figure~\ref{fig:k-c-w_initial} plots \(k_{\rm cubic}\) with respect to \(c_0\) and \(\omega\), while maintaining \(\kappa\) and \(\epsilon_{\rm misfit}\) constant at 1.258 n-J/m (or 10$^{-9}$ $\cross$ J/m) and 0.62\%, respectively. Figure~\ref{fig:k-k-eps_initial} plots \(k_{\rm cubic}\) with respect to \(\kappa\) and \(\epsilon_{\rm misfit}\), while maintaining \(c_0\) and \(\omega\) at 0.831 and 0.256 G-J/m$^3$ (or 10$^9$ $\cross$ J/m$^3$) respectively. The standard deviation of the model is displayed by the colors in the 3D surface plots. We observe a strong dependence of the precipitate shape parameter $k_{\rm cubic}$ on the double-well barrier height $\omega$. There is a higher uncertainty from the middle to upper bound of the \(\omega\) values. Figure~\ref{fig:k_final} displays the surrogate model for $k_{\rm cubic}$ after the final iteration of the GPR model and after including all the new samples suggested with BO. The high uncertainty regions are now located at the edges of the parameter space, which is preferred.  
The final model also agrees qualitatively with the initial model in that there is a strong dependence of precipitate morphology with respect to \(\omega\). The other three phase-field parameters ($c_0$, $\kappa$, and $\epsilon_{\rm misfit}$) have little influence on the precipitate morphology. Both initial and final models predicts similar trends between the precipitate size and the four phase-field parameters, \blue{but} the predicted values are significantly changed after the addition of new samples. This is better observed between Figures.~\ref{fig:k-k-eps_initial} and ~\ref{fig:k-k-eps_final}, where the predicted $k_{\rm cubic}$ value changes from approximately 0.50 to 0.91. Figure~\ref{fig:k-w} compresses the multi-dimensional model to a 2-D plot, which depicts the relationship between the most sensitive parameter, \(\omega\), and the precipitate morphology parameter, \(k_{\rm cubic}\).

\begin{figure}
    \centering
    \includegraphics[width=0.7\textwidth]{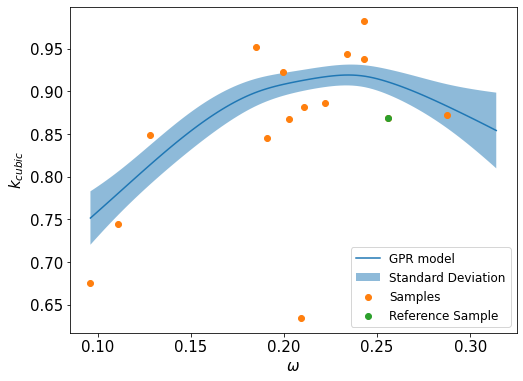}
    \caption{Final GPR model of precipitate morphology versus the double well barrier potential. The green dot is the reference sample and the orange dots are all the samples. The blue solid line marks the GPR model prediction and the light blue shade indicates the standard deviation, which displays a strong dependence of the $\gamma^\prime$(L1$_2$) precipitate morphology on the double-well barrier height \(\omega\).}
    \label{fig:k-w}
\end{figure}

\section{Discussion}
From the GPR model of the precipitate size as function of input parameters, we determined that the initial Co concentration is the dominant parameter influencing the equilibrium precipitate size. As displayed in Fig.~\ref{fig:r-c}, the GPR model fits all the sample points relatively well, despite having different \(\omega\), \(\kappa\), and \(\epsilon_{\rm misfit}\) values. The precipitate size is predicted to be smaller with increasing Co concentration in the ternary Co-Al-W alloy system. This result is consistent with experimental observations \cite{lass2018}. Since Al is a constant across the two phases, the increase in Co concentration implies a decrease in the W concentration in the system. We have observed that W partitions towards the $\gamma^\prime$(L1$_2$) phase and stabilizes it, which explains how the equilibrium precipitate size is smaller when there is a higher Co concentration \cite{chung2020}. The results also imply that the precipitate size is heavily dependent on the chemical free energy of the system, since the alloy composition is a major component of the formulation of the chemical free energy. The slight downward curve where the initial concentration is smaller than 0.81 is most likely due to a lack of data at the edge of the search space, where the \(\gamma'\)(L1$_2$) precipitates are close to being unstable. 

From the precipitate morphology GPR model, we observe that the phase field double-well barrier height \(\omega\) strongly influences the $\gamma^\prime$(L1$_2$) precipitate morphology. Figure~\ref{fig:k-w} shows a compressed 2-D plot of the $\gamma^\prime$(L1$_2$) precipitate morphology and \(\omega\). There appears to be an upward trend in \(k_{\rm cubic}\) with increasing \(\omega\). The downward trend of \(k_{\rm cubic}\) at high \(\omega\) values may be due to a lack of data points in that region. The precipitate morphology is determined by the balance between the interfacial free energy and elastic energy \cite{thompson1999}. The increase in \(\omega\) leads to an increase in interfacial free energy between the \(\gamma\)(f.c.c.) and $\gamma^\prime$(L1$_2$) phase. Also, the double-well barrier height, (\(\omega\)), controls how much the energy barrier is needed to be overcome to a phase change. The results suggest that a larger barrier leads to a more cubic shaped $\gamma^\prime$(L1$_2$) precipitate. 

The precipitate size and morphology models are plotted with other samples as reference points to verify that our observations are consistent at other regions of the parameter space. Refer to the Appendix for the GPR models plotted with Samples 11 and Sample 18 as reference points. 
The additional GPR model surface plots with different reference points are like Figures. \ref{fig:r_final} and \ref{fig:k_final}. Similarly, the precipitate size and morphology are most sensitive to the initial Co concentration $c_0$ and double-well barrier height $\omega$ respectively.

It is interesting that the general surface response trend for both $\gamma^\prime$(L1$_2$) precipitate size and morphology GPR models with the initial five samples is consistent with the final iteration of the model with a full set of samples. The additional data points help to improve the accuracy of the model, which is better observed in the precipitate size GPR model, shown in Figures.~\ref{fig:r_initial} and \ref{fig:r_final}.

Compared to the previous phase-field sensitivity analysis\cite{wu2022}, we observe that the dominant factor for precipitate size is its initial concentration and the precipitate morphology is the cross term of the double well potential and lattice parameter misfit strain. The observation for the precipitate size model in this study agrees very well with the sensitivity analysis results obtained by Wu \emph{et al.}\cite{wu2022}. 

This further proves that the initial composition of the Co-Al-W ternary alloys is the most important factor to consider when modifying the $\gamma^\prime$(L1$_2$) precipitate's size. Alternatively, the precipitate morphology GPR model predicts that the double-well barrier height \(\omega\) dominates the precipitate's morphological evolution. A possible explanation for this difference could originate from how the two models predicts differently. The bi-linear fit predicts normalized parameters, which limits the range of individual parameters. For parameters that have a smaller range, a small difference could appear to be very sensitive. If we only consider the single terms in the bi-linear fit sensitivity analysis, the double well potential (\(\omega\)) is the parameter that has the strongest influence to the precipitate morphology.  The general positive trend observed in Fig. \ref{fig:k-w} also agrees with the positive sensitivity coefficient previously calculated \cite{wu2022}.

By comparing Figures.~\ref{fig:r-c} and \ref{fig:k-w}, we conclude that the precipitate morphology model is less accurate compared to the precipitate size model. The precipitate morphology GPR model requires a higher noise tolerance to avoid over-fitting the data. Without a higher noise tolerance factor, the GPR model response surface will fluctuate significantly to accommodate the noisy data, which is less useful for interpreting the GPR model.
A possible explanation for the noisy dataset could be from the representation of precipitate morphology, Eq. \ref{k_cubic_eqn}. This equation expresses how cubic the precipitate is based on the ratio between the maximum and minimum distance to the interface of the $\gamma^\prime$(L1$_2$) precipitate. Some of the simulated precipitates, however, have an irregular morphology, such as having slightly concave sides on the surfaces of the cuboidal precipitates. These irregularities translate into the noise in the dataset. In this case, the model can only provide a qualitative understanding of how the parameters influence the equilibrium $\gamma^\prime$(L1$_2$) precipitate morphology. A possible way to improve the model is to use a more precise description of precipitate morphology, which includes any irregular features. 

The GPR model is a more sophisticated data-driven approach in fitting a surrogate model compared to a standard bi-linear fit. This method fits a probability distribution over many functions and the predicted output is generated with a maximum likelihood estimation. This can be used to generate a surface response that is interpretable, which is useful for sensitivity analyses. The GPR model is also considered a great surrogate model for situations where there is a small data set. Since the GPR model generates a probability distribution over functions, we can easily obtain the standard deviation of the model as the uncertainty. Uncertainty quantification is also required for Bayesian Optimization, which is a more efficient method in selecting new samples to update the surrogate model iteratively compared to a random selection or grid search method. The BO approach is beneficial for scenarios where measuring a data point is expensive and it is desirable to minimize the number of calculations or measurements needed to update the model. The 3-D phase-field simulations computed in this study can take up to one week of computation time with 144 cores. This approach balances exploration and exploitation to determine new data points that are optimized for an objective that is still significant for the model. By selecting the expected improvement equation, the objective is to minimize the uncertainty of the model. 
An exploration term is added to avoid the edges of the exploration space, which is less important for interpreting the surface response of the GPR model. 

\section{Conclusions}
In this study, a phase-field sensitivity analysis is conducted by generating GPR models for the $\gamma^\prime$(L1$_2$) precipitate size and morphology. The initial samples are selected with Latin hypercube, which maximizes the distance between each point within the parameter space. New samples are determined with the Bayesian Optimization algorithm that targets high uncertainty regions to update the surrogate models. The precipitate size and morphology GPR models predicted that the dominant phase-field parameter is the initial composition and double well barrier potential, respectively. The dominant phase-field parameter predicted by the GPR models agrees well with the previously identified dominant parameter from the standard bi-linear fit. 

The precipitate size model has a high accuracy due to its precise description. Alternatively, the precipitate morphology description does not include small variations in a precipitate's morphology, which translates into noise in the dataset. This results in a GPR model that requires higher noise tolerance and lower accuracy. Therefore, the precipitate size model may be able to provide a quantitative relation between parameters, but the precipitate morphology model can only provide a qualitative understanding at this current iteration.

\section*{Acknowledgments}
This work was performed under financial assistance award 70NANB19H005 from the U.S. Department of Commerce, National Institute of Standards and Technology as part of the Center for Hierarchical Material Design (CHiMaD).  We gratefully acknowledge the computing resources provided on Bebop and Blues, which are high-performance computing clusters operated by the Laboratory Computing Resource Center at Argonne National Laboratory. We particularly thank the participants of the Phase Field Workshops held in Evanston, IL, to whom we are deeply indebted for invaluable feedback and comments.   

\bibliographystyle{plain}
\bibliography{BO_bibliography.bib}

\newpage

\section{Appendix}
The figures in this section plots the same precipitate size and morphology GPR models with different reference samples. The observations presented in the Discussion section can be similarly observed in these figures. This supports the conclusion that the $\gamma'$(L1$_2$) precipitate size is most sensitive to the initial Co concentration $c_0$ and the $\gamma'$(L1$_2$) precipitate morphology is most sensitive to the double-well potential height $\omega$.

\begin{figure}[h]
\renewcommand{\thefigure}{A\arabic{figure}}
\setcounter{figure}{0}
\begin{subfigure}{0.45\textwidth}
    \centering
    \includegraphics[width=\textwidth]{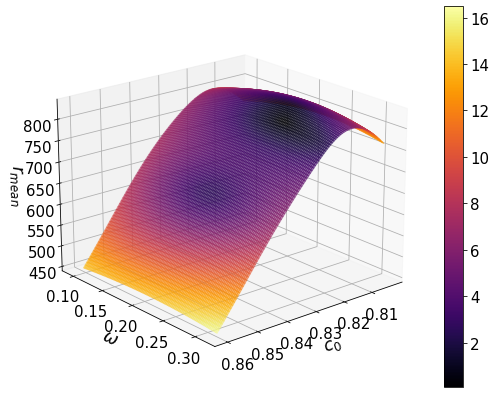}
    \caption{}
    \label{fig:new2_r-c-w}
\end{subfigure}
\begin{subfigure}{.45\textwidth}
    \centering
    \includegraphics[width=\textwidth]{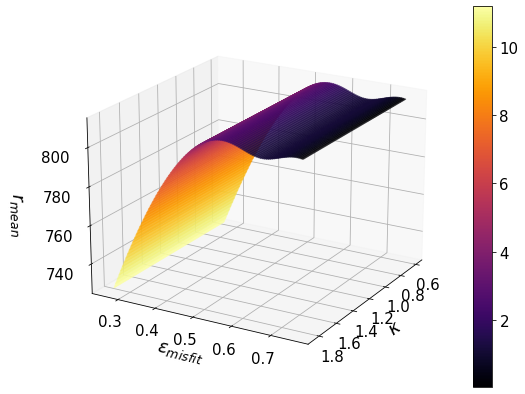}
    \caption{}
    \label{fig:new2_r-k-eps}
\end{subfigure}
\caption{Final GPR surrogate model of precipitate size fitted with all samples with Sample 11 as the reference point. The surface plot color and the scale bar to the right represents the standard deviation of the model. \(r_{\rm mean}\) and standard deviation are in nm and \(c_0\) and \(\epsilon_{\rm misfit}\) are dimensionless, \(\omega\) is G$\cross$J/m$^3$ (or 10$^9$ $\cross$ J/m$^3$), and \(\kappa\) is n$\cross$J/m (or 10$^{-9}$ $\cross$ J/m). (a) is plotted with $\kappa$ = 1.007 n$\cross$J/m (or 10$^{-9}$ $\cross$ J/m) and $\epsilon_{\rm misfit}$ = 0.753\%. (b) is plotted with $c_0$ = 0.8216 and $\omega$ = 0.243 G$\cross$J/m$^3$ (or 10$^9$ $\cross$ J/m$^3$). The plots have different z-axis scaling. This demonstrates that the precipitate size is most sensitive to the initial Co concentration $c_0$.}
\label{fig:new2_r}
\end{figure}

\begin{figure}
\renewcommand{\thefigure}{A\arabic{figure}}
\begin{subfigure}{0.45\textwidth}
    \centering
    \includegraphics[width=\textwidth]{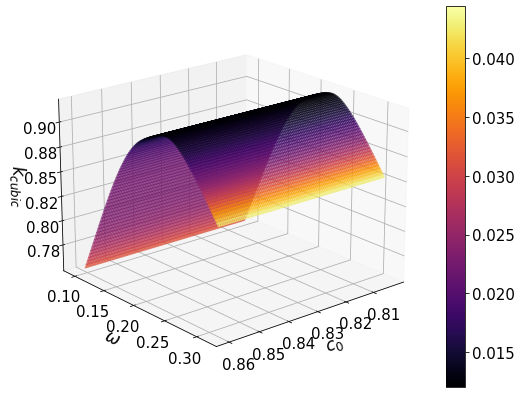}
    \caption{}
    \label{fig:new2_k-c-w}
\end{subfigure}
\begin{subfigure}{.45\textwidth}
    \centering
    \includegraphics[width=\textwidth]{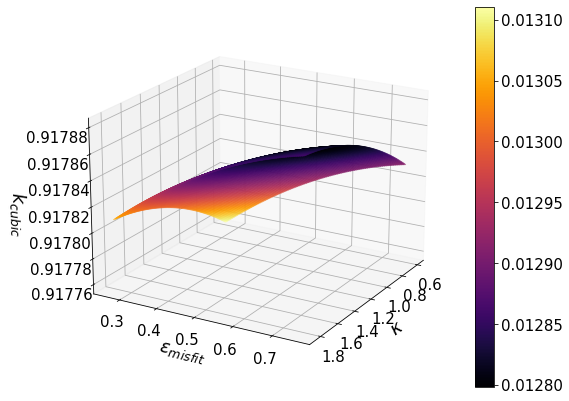}
    \caption{}
    \label{fig:new2_k-k-eps}
\end{subfigure}
\caption{Final GPR surrogate model of precipitate morphology fitted with all samples with Sample 11 as the reference point. The surface plot color and the scale bar to the right represents the standard deviation of the model. $k_{\rm cubic}$, standard deviation, \(c_0\), and \(\epsilon_{\rm misfit}\) are dimensionless, \(\omega\) is G$\cross$J/m$^3$ (or 10$^9$ $\cross$ J/m$^3$), and \(\kappa\) is n$\cross$J/m (or 10$^{-9}$ $\cross$ J/m). (a) is plotted with $\kappa$ = 1.007 n$\cross$J/m (or 10$^{-9}$ $\cross$ J/m) and $\epsilon_{\rm misfit}$ = 0.753\%. (b) is plotted with $c_0$ = 0.8216 and $\omega$ = 0.243 G$\cross$J/m$^3$ (or 10$^9$ $\cross$ J/m$^3$). The plots have different z-axis scaling. This demonstrates that the precipitate morphology is most sensitive to the double-well barrier height $\omega$.}
\label{fig:new2_k}
\end{figure}

\begin{figure}
\renewcommand{\thefigure}{A\arabic{figure}}
\begin{subfigure}{0.45\textwidth}
    \centering
    \includegraphics[width=\textwidth]{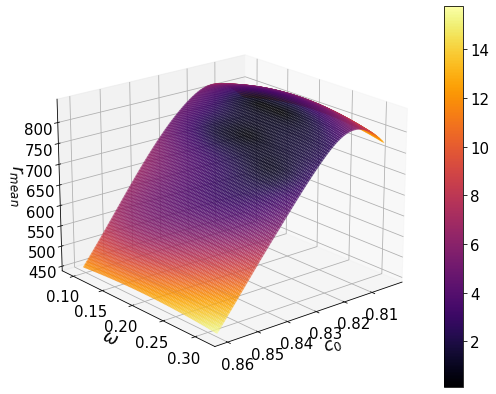}
    \caption{}
    \label{fig:new9_r-c-w}
\end{subfigure}
\begin{subfigure}{.45\textwidth}
    \centering
    \includegraphics[width=\textwidth]{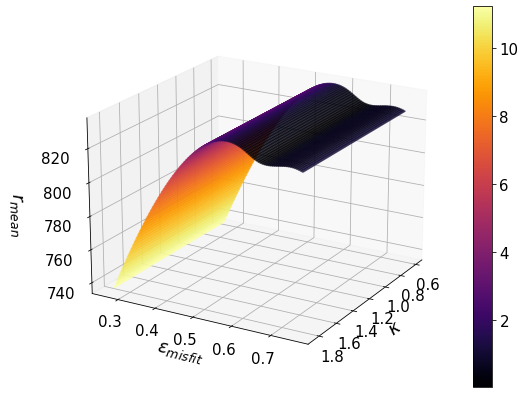}
    \caption{}
    \label{fig:new9_r-k-eps}
\end{subfigure}
\caption{Final GPR surrogate model of precipitate size fitted with all samples with Sample 18 as the reference point. The surface plot color and the scale bar to the right represents the standard deviation of the model. \(r_{\rm mean}\) and standard deviation are in nm and \(c_0\) and \(\epsilon_{\rm misfit}\) are dimensionless, \(\omega\) is G$\cross$J/m$^3$ (or 10$^9$ $\cross$ J/m$^3$), and \(\kappa\) is n$\cross$J/m (or 10$^{-9}$ $\cross$ J/m). (a) is plotted by keeping $\kappa$ = 1.382 n$\cross$J/m (or 10$^{-9}$ $\cross$ J/m) and $\epsilon_{\rm misfit}$ = 0.646\%. (b) is plotted by keeping $c_0$ = 0.8189 and $\omega$ = 0.2222 G$\cross$J/m$^3$ (or 10$^9$ $\cross$ J/m$^3$). The plots have different z-axis scaling. This demonstrates that the precipitate size is most sensitive to the initial Co concentration $c_0$.}
\label{fig:new9_r}
\end{figure}

\begin{figure}
\renewcommand{\thefigure}{A\arabic{figure}}
\begin{subfigure}{0.45\textwidth}
    \centering
    \includegraphics[width=\textwidth]{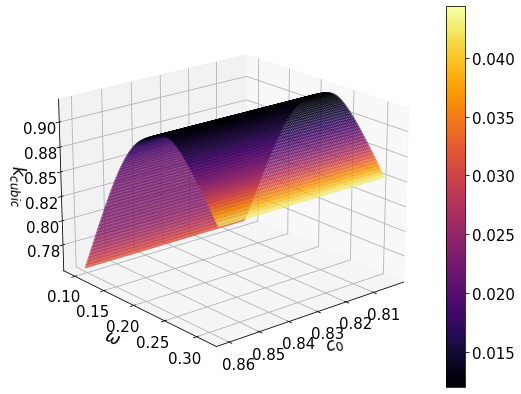}
    \caption{}
    \label{fig:new9_k-c-w}
\end{subfigure}
\begin{subfigure}{.45\textwidth}
    \centering
    \includegraphics[width=\textwidth]{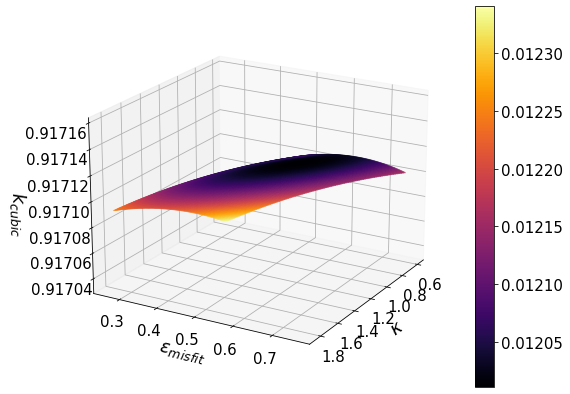}
    \caption{}
    \label{fig:new9_k-k-eps}
\end{subfigure}
\caption{Final GPR surrogate model of precipitate morphology fitted with all samples with Sample 18 as the reference point. The surface plot color and the scale bar to the right represents the standard deviation of the model. $k_{\rm cubic}$, standard deviation, \(c_0\), and \(\epsilon_{\rm misfit}\) are dimensionless, \(\omega\) is G$\cross$J/m$^3$ (or 10$^9$ $\cross$ J/m$^3$), and \(\kappa\) is n$\cross$J/m (or 10$^{-9}$ $\cross$ J/m). (a) is plotted by keeping $\kappa$ = 1.382 n$\cross$J/m (or 10$^{-9}$ $\cross$ J/m) and $\epsilon_{\rm misfit}$ = 0.646\%. (b) is plotted by keeping $c_0$ = 0.8189 and $\omega$ = 0.2222 G$\cross$J/m$^3$ (or 10$^9$ $\cross$ J/m$^3$). The plots have different z-axis scaling. This demonstrates that the precipitate morphology is most sensitive to the double-well barrier height $\omega$.}
\label{fig:new9_k}
\end{figure}

\clearpage

\section{Supplementary material: Clarification on units}
The units for interfacial energy are energy per unit area. In this article, we employ J/m$^2$, where J is joule and m is meter. To keep the other parameters consistent with this convention, they are labeled as listed in Table S\ref{units_table}.

\begin{table} [h]
\renewcommand\tablename{S}
\begin{center} 
    \caption{\label{units_table}The convention for interfacial energy related to parameters employed in this article.}
    \begin{tabular}{ c c c c }
    \hline
    Parameter & Parameter Name & Unit & SI Unit Equivalent \\
    \hline
    $\omega$ & Double-well barrier height & G$\cross$J/m$^3$ & 10$^9$ $\cross$ J/m$^3$ \\
    $\kappa$ & Gradient energy density coefficient & n$\cross$J/m & 10$^{-9}$ $\cross$ J/m \\
    $L$ & Kinetic coefficient & 10$^{-45}$ $\cross$m$^3$s$\cdot$J & 10$^{-45}$ $\cross$ m$^3$s$\cdot$J \\
    \hline
    \end{tabular}
\end{center}
\end{table}

\end{document}